\documentclass[10pt,journal,compsoc]{IEEEtran}
\usepackage{blindtext}

\usepackage{amsmath,amssymb,amsfonts}
\usepackage{enumitem}
\usepackage{algorithmic}
\usepackage{graphicx}
\usepackage{textcomp}

\usepackage[colorlinks=true,citecolor=blue]{hyperref}

\ifCLASSOPTIONcompsoc
   \usepackage[nocompress]{cite}
\else
    \usepackage{cite}
\fi

\ifCLASSINFOpdf
 
\else
 
\fi

\usepackage{graphicx}
\usepackage{xcolor}

\begin{document}

\title{A Systematic Literature Review on the security of Unmanned Aerial Vehicle System}

\author{\IEEEauthorblockN{Tirth Patel \IEEEauthorrefmark{1}  }\\
\IEEEauthorblockA{\textit{\IEEEauthorrefmark{1}School of Engineering,} \\
\textit{University of Guelph, Ontario, Canada  }\\
tpatel04@uoguelph.ca}
\\
 \and
\IEEEauthorblockN{ Niyatiben Salot \IEEEauthorrefmark{2}}\\
\IEEEauthorblockA{\textit{\IEEEauthorrefmark{2}School of Engineering,} \\
\textit{University of Guelph, Ontario, Canada }\\
nsalot@uoguelph.ca}
\\
 \and
\IEEEauthorblockN{ Vrusha Parikh \IEEEauthorrefmark{3}}\\
\IEEEauthorblockA{\textit{\IEEEauthorrefmark{3}School of Engineering,} \\
\textit{University of Guelph, Ontario, Canada }\\
parikhv@uoguelph.ca}}




\IEEEtitleabstractindextext{%
\begin{abstract}
\textcolor{black}{{Unmanned aerial vehicles (UAVs) are becoming more common, and their operational range is expanding tremendously, making the security aspect of the inquiry essential. This study does a thorough assessment of the literature to determine the most common cyberattacks and the effects they have on UAV assaults on civilian targets. The STRIDE assault paradigm, the challenge they present, and the proper tools for the attack are used to categorize the cyber dangers discussed in this paper. Spoofing and denial of service assaults are the most prevalent types of UAV cyberattacks and have the best results. No attack style demands the employment of a hard-to-reach gadget, indicating that the security environment currently necessitates improvements to UAV use in civilian applications.}}
\end{abstract}

}

\maketitle

\IEEEdisplaynontitleabstractindextext

\IEEEpeerreviewmaketitle

https://www.overleaf.com/project/632b2e04626bd814009a26b6

\section{Introduction}\label{sec:introduction}
The Unmanned Aerial Vehicle is an acronym for UAV, an air vehicle without any onboard pilot. The aircraft is controlled remotely by a controller which is on the ground or the UAV can fly autonomously with the complex pre-programmed model \cite{hallermann2014visual}. It is also called a Remotely Piloted Vehicle (RPV) when pilotless aircraft are operated from the ground via electronically controlled equipment. UAVs are used for observatory and tactic planning. UAVs are capable of maintaining the flight level which is powered by jet propellers or reciprocating engines. UAVs can be categorized into use for government, commercial and hobby purposes.\cite{liu2020unmanned} There are a plethora of functional uses for UAVs like military training by providing both sky and ground coverage of arms with a specific target and also helping in simulating enemies' air vehicles, reconnaissance, combating, for logistics: delivering cargo, improving technologies with research and development in UAVs and commercial/civil purpose of taking agriculture information and aerial photography. Unmanned Aerial vehicles have a great use to a certain level easing the work of humankind with economical financial cost\cite{lv2019security}.
\\Modern-day UAVs are helpful for tracking wildfires, observing the evolution of the hurricane, a survey of landscaping, and disaster relief. Several regions worldwide have put in a joint effort with innovators and have discovered ingenious ways of use\cite{chamola2020comprehensive}. A wireless network controls the device in the air\cite{lv2019security}. The directions given to the drone and the data collected must be supported via a network\cite{aftab2019hybrid,lv2019ieee}. Encryption of the data collected through UAV must be checked for the security performance of the network\cite{zhang2018guest}. The data collected through the drones vary on the purpose of the application and the device used. Data collected for agriculture has information on humidity, dampness of the soil, etc. while the data collected from the health department has information on heartbeats, disease type, etc. Data collected is not always private and confidential but data obtained should be secured when transmitted with the network and further as well\cite{alladi2020consumer}.
\\Due to the escalating worries over the security of UAV software, there has been very little exploratory work. The existing drone security environment has been thoroughly examined as part of this research\cite{parlin2018jamming}.\cite{ly2021cybersecurity} The Spoofing, Tampering, Repudiation, Information Disclosure, Denial of Service, and Elevation of Privilege (STRIDE) threat model is used to categorize the cyberattacks in the research and identify the dangers they represent as well as the tools required to carry them out.

\begin{figure}[h]
	    \centering
	    \includegraphics[width=0.9\linewidth,keepaspectratio]{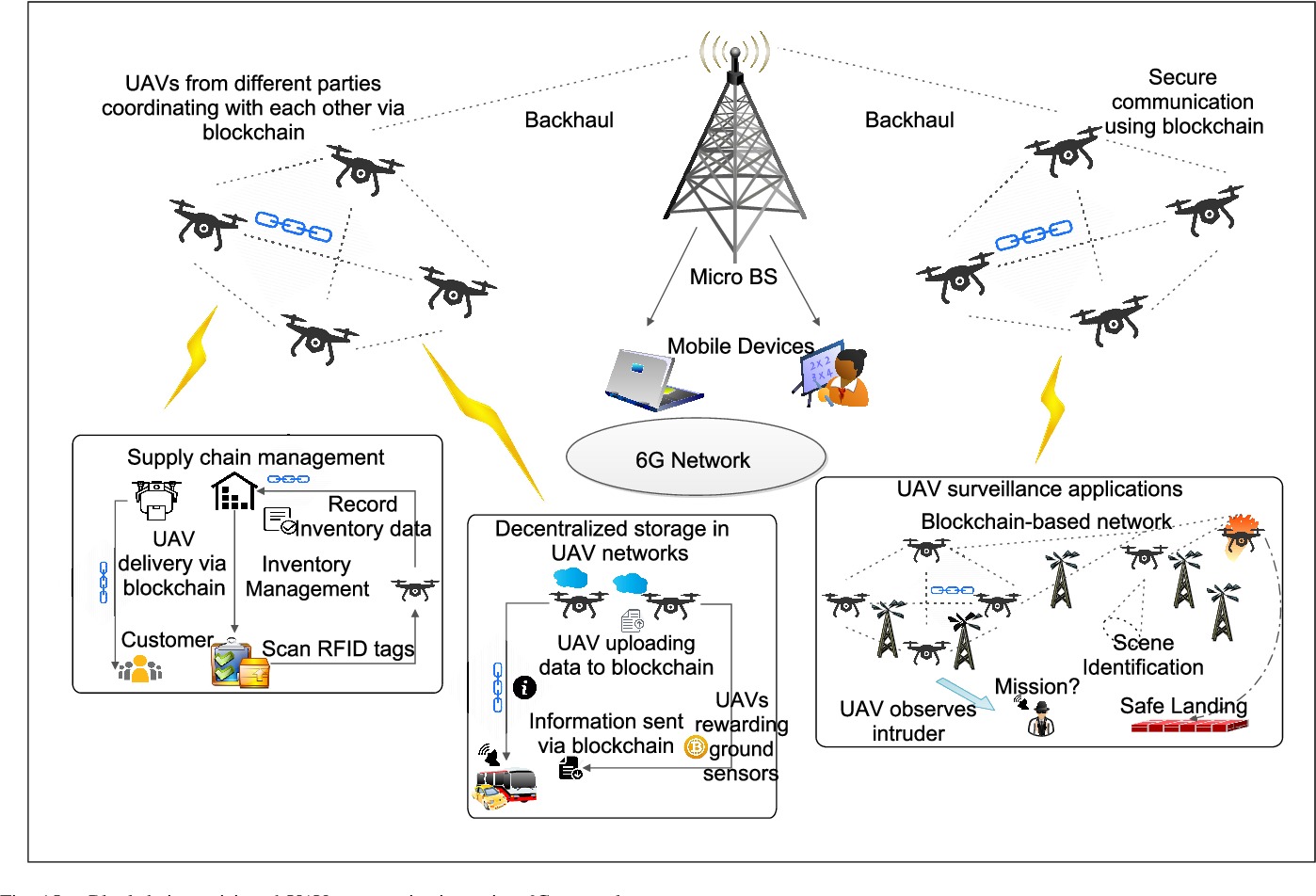}
	    \caption{Drones Applications}
	    \label{MyFig1}
\end{figure}

\subsection{Prior research}
Unmanned aircraft are seen as a new and developing category of "flying IoT" gadgets \cite{shakhatreh2019unmanned}. They include a number of applications.
As stated by Alladi et al.\cite{lv2019security} Unmanned Aerial Vehicle is more susceptible to being lost, hijacked, or destroyed since it is deployed in open-air space. Many network problems need to be taken into consideration like intra-communication, surveillance, protection of data in the air, data storage, and management because drone technology is now omnivorous. Blockchain, for instance, is a distributed ledger that uses encryption techniques like the hash function and public keys cryptography to protect shared data\cite{alladi2020applications}. Additionally, it may be used to confirm the accuracy of the data that has been processed and to increase the safety and openness of UAVs. Flaws in security systems have many severe implications.
\\However, the upsurge in manufacturing to meet the rising demand has created a number of security flaws or vulnerabilities\cite{gudla2018defense}. A number of issues make it feasible for an enemy to crash the device, causing the drone to malfunction and hurting network users. Significant flaws allow the hacker to totally control the control system and seize the UAV or anything it is holding\cite{westerlund2019drone}. Drones are often deployed for recording or monitoring in both private and commercial settings, and any cyberattack may frequently entail worries about privacy and other hazards. 
\\\textbf{Challenges of drones as security and privacy:} Shakhatreh et al. \cite{shakhatreh2019unmanned} discussed the civil uses of Drones and their main difficulties. A study of the UAVs' cybersecurity vulnerabilities was done by Krishna et al. \cite{krishna2017review}. The authors suggested a taxonomy to categorize various UAV cyberattack types.
\\\textbf{Privacy and Security issues in the communication of drones:} The significant security concerns of Aircraft wireless connectivity were reviewed by Fotouhi et l. \cite{fotouhi2019survey}. According to Mishra et al. \cite{mishra2020survey}, the incorporation of Drones into cellular networks like 5G introduces security problems that must be adequately addressed and analyzed by the scientific community. Hayat and co. addressed the UAV's privacy, security, and safety concerns networks from the perspective of communication. then offered UAV networks' standard communication requirements for a UAV deployment that is safe, secure, and protects privacy.

\subsection{Research Goals}
The objective of this research is to review earlier research, synthesize its conclusions, and focus on the security of Unmanned Aerial vehicles. To make the work more focused, we developed three research questions, as shown in Table. \ref{Mytable1}.
\begin{table}[h!]
\caption{Research questions.} 
\label{Mytable1}
\setlength{\tabcolsep}{3pt}
\begin{tabular}{p{106pt} p{109pt} }
\hline
 Research Questions(RQ) & Discussion\\
 \hline
 \vspace{0.15 mm} \textbf{RQ1:} What are recent advances for pre-venting information loss of UAVs? 
  & \vspace{0.15 mm}Routing protocols are necessary for multi-UAV networks to offer dependable end-to-end data transfer between UAV nodes. In the literature, a number of network applications have been presented with various classifications. One method divides these protocols into two categories: data forwarding and network architecture.
\\
 \vspace{0.15 mm}  \textbf{RQ2:} Is any secured method available for communications the drones?
 & \vspace{0.15 mm}Communication methods vary from the use case of drones, thus we need a standard method for communication. In the papers, there are several methods proposed that are being used and can be used to mitigate the risk and have smooth communication. \\
 \vspace{0.15 mm}  \textbf{RQ3:} How is the security measured for overtaking attacks done at various types of stages like level of communication, hardware level, sensor level, and software level?
 & \vspace{0.15 mm} These stages are well structured for the best use of UAVs. In the papers, a very small number of studies have talked about all of the stages. We found that most of the studies have a level of communication and a few have hardware-level security measurements.\\
 \hline
\end{tabular}
\end{table}
\subsection{Contribution and Layouts}
Current research is complemented by this SLR, which offers the For those interested in blockchain technology, the following contributions using cyber security to advance their work.
\begin{itemize}
\item We identify 30 primary studies during the search of IoT and UAV papers in the specific field.
\item We analyze the data collected by the various studies and present the findings to provide an updated view of the UAV. Through the review, we hope to gain a deeper understanding of this subject.
\item The goal of this review is to examine the several strategies that can be used to increase the safety of various cyber technologies.
\item To encourage additional research in this field, we establish guidelines and make representations.
\end{itemize}

Table \ref{Mytable1} summarizes the three relevant research questions and provides a discussion section where relevant topics can be discussed. The discussion section will also cover the various aspects of UAV Security.

\section{Research Methodology}
There are many studies have been carried out on Unmanned aerial vehicles. Rather than its speed, technology, safety, or its design. For the systematic review, we have taken guidance from the paper of Kitchen ham and Charters\cite{kitchenham2007guidelines}. We have agreed to go through phases of Selection, Scanning, Prioritizing, and Extraction to get the best for our SLR.

\subsection{Selection of primary studies}
Research studies were found by passing the search keyword or string to the search engines or publications to get the primary studies. The keyword selection was done to proliferate research and can able to find the paper for answering the research questions. The search string was as below:
\\The data was extracted using the boolean operator AND and OR, which were used as search strings:
\\The boolean operators are as follows for IEEE: \textit{“Unmanned Aerial Vehicle” OR "IoT" AND (“ Cyber Security” OR “security”)}.
\\Based on INSPEC the search strings are:\textit{ “IoT” AND “UAV” AND “ communication” AND (“networking”  OR “security”)}.

\vspace{1.5 mm}The digital library platforms used during the research are as follows:
\\- IEEE Xplore Digital Library 
\\- SpringerLink 
\\- Google Scholar
\\- UOG Library
\\- ScienceDirect
\\The search was run through description, abstract, or title on the basis of the platforms. This search was done on 1st Oct 2022. After filtering from Inclusion and Exclusion criteria, which will be discussed further. The criterion enabled us to generate a set of outcomes that were then subjected to Wohlin's snowballing process\cite{wohlin2014guidelines}.

\subsection{Inclusion and Exclusion criteria}
The goal of this article is to provide a review of the literature on cybersecurity in UAVs, with a particular focus on cybersecurity issues and solutions in this setting. The following research questions were posed for this study:
\\RQ1: What are recent advances for preventing information loss of UAVs? 
\\RQ2:  Is any secured method available for communications the drones?
\\RQ3: How is the security measured for overtaking attacks done at various types of stages like level of communication, hardware level, sensor level, and software level?
\\The study selection criteria were used to determine the most relevant papers for the review. The results of the search were then analyzed using Google scholar to find the most relevant papers. Table \ref{Mytable2}  illustrates the criteria for the inclusion and exclusion of the papers for the primary studies.
\subsection{Selection results}
Hundreds of research papers were searched under the above-mentioned search string related to UAV and Cyber Security. After removing duplicates and irrelevant studies, 450 research papers were there to analyze using Inclusion and exclusion criteria. After the criteria were applied a total of 25 papers remained. Snowballing both forward and backward revealed an additional 2 and 3 papers, bringing the total number of papers to be included in this SLR to 30

\subsection{ Quality assessment }
In this section, we assess the research papers on the basis of their quality of writing, their ideas, and their experimental ideas based on guidance by Kitchenham and Charter\cite{kitchenham2007guidelines}. This assessment gives us knowledge of the content of the primary studies. This assessment is followed by the assessment described by Hosseini et al.\cite{8097045}. Randomly 4-5 papers will be tested on 4 stages of assessment, in each stage, we look for specific data, if the research paper passes all the stages, we would include those papers in our SLR, otherwise, we have to drop them. Those stages are as follows.
After this assessment, it has been detected that there are 9 studies that do not meet the requirement, that's why removed from the SLR. Refer to table.

\begin{table}[h!]
\caption{Inclusion and exclusion for primary studies} 
\label{Mytable2}
\setlength{\tabcolsep}{3pt}
\begin{tabular}{p{106pt} p{109pt} }
\hline
 Criteria for Inclusion & Criteria for exclusion\\
\hline
 \vspace{0.15 mm}The paper must have information regarding UAVs. &  \vspace{0.15 mm}Paper focusing mainly on the economic side of the UAV.\\
 \vspace{0.15 mm} The paper must have cyber security or UAV security-related data. &  \vspace{0.15 mm}Paper with just cyber security. \\
 \vspace{0.15 mm}The paper must possess information regarding cyber attack and their prevention. &  \vspace{0.15 mm} Any global blogs or internet source data. \\
\hline
\end{tabular}
\end{table}

The stages of finding the excluded studies, as seen in Table \ref{Mytable3}, are as follows:
\\Stage 1: \textbf{Unmanned Aerial Vehicle.} The studies must be on UAVs and their functions or have a case study on UAVs.
\\Stage 2:\textbf{Security.} The studies must have information on the security of UAVs.
\\Stage 3: \textbf{Cyber Attacks.} Studies must have data on the cyber attack that can be done on drones
\\Stage 4: \textbf{Data Context} The studies must have ample information on each stage and can be verified.

\subsection{ Data extraction }

Now that we have our final research papers, we plan how to extract information from them. As our topic is related to security, the first thing we look for is security purposes. We want those papers, which has in-depth data regarding the security of the UAVs and we will use data on UAVs provided by other papers to examine it . We created two categories in which we divided the papers to conduct further research. Those categories are:
\\UAV: Use-case and Functions.
\\Security: Threats, attacks and possible solutions.
\subsection{Data analysis}
We gathered the information included in the qualitative and quantitative data categories in order to achieve the goal of responding to the study questions. Additionally, we performed a meta-analysis on the studies that underwent the last step of data extraction.

\subsubsection{Publications over time}
The concept of the Unmanned Aerial Vehicle goes back to 1849. After that many researchers studied this concept and presented their ideas on this. However, there is a very inclination in field research after 2010. In 2010 there is 200 studies presented on Unmanned Aerial vehicles. As we see in the chart, there is the most upward trend in the use case and research of Unmanned Aerial Vehicle over the period of time.

\begin{table}[h!]
\caption{Excluded studies} 
\label{Mytable3}
\setlength{\tabcolsep}{3pt}
\begin{tabular}{p{108pt} p{107pt} }
\hline
Checklist for the Criteria Stages& Excluded Studies 
\\
 \hline
 \vspace{0.15 mm}Stage 1: Unmanned Aerial Vehicles.  & \vspace{0.15 mm}\cite{thakur2015investigation},\cite{waharte2010supporting}
 \\Stage 2: Security & \vspace{0.15 mm} \cite{bortoff2000path},\cite{cremer2022cyber} \\
 \vspace{0.15 mm}\\Stage 3: Cyber Attacks & \vspace{0.15 mm} \cite{waharte2010supporting},\cite{mozaffari2019tutorial},\cite{bortoff2000path}\\
 \vspace{0.15 mm}Stage 4: Data Context. & \vspace{0.15 mm} \cite{basan2021self}, \cite{KONERT2021292}\\
 
 \hline
\end{tabular}

\end{table}

\begin{figure}[h]
	    \centering
	    \includegraphics[width=0.9\linewidth,keepaspectratio]{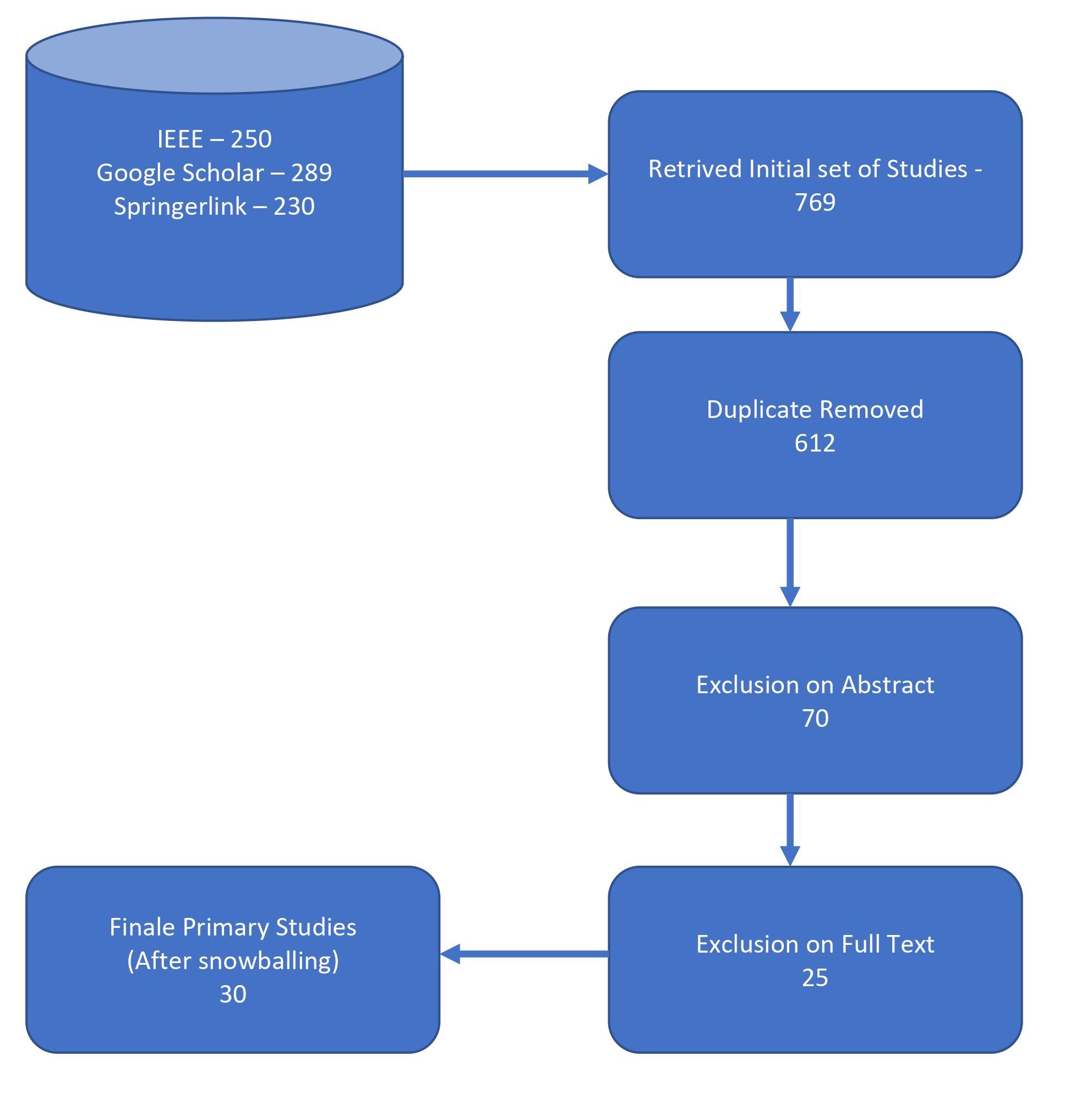}
	    \caption{Attrition of papers through processing}
	    \label{MyFig2}
\end{figure}

\begin{figure}[h]
	    \centering
	    \includegraphics[width=0.9\linewidth,keepaspectratio]{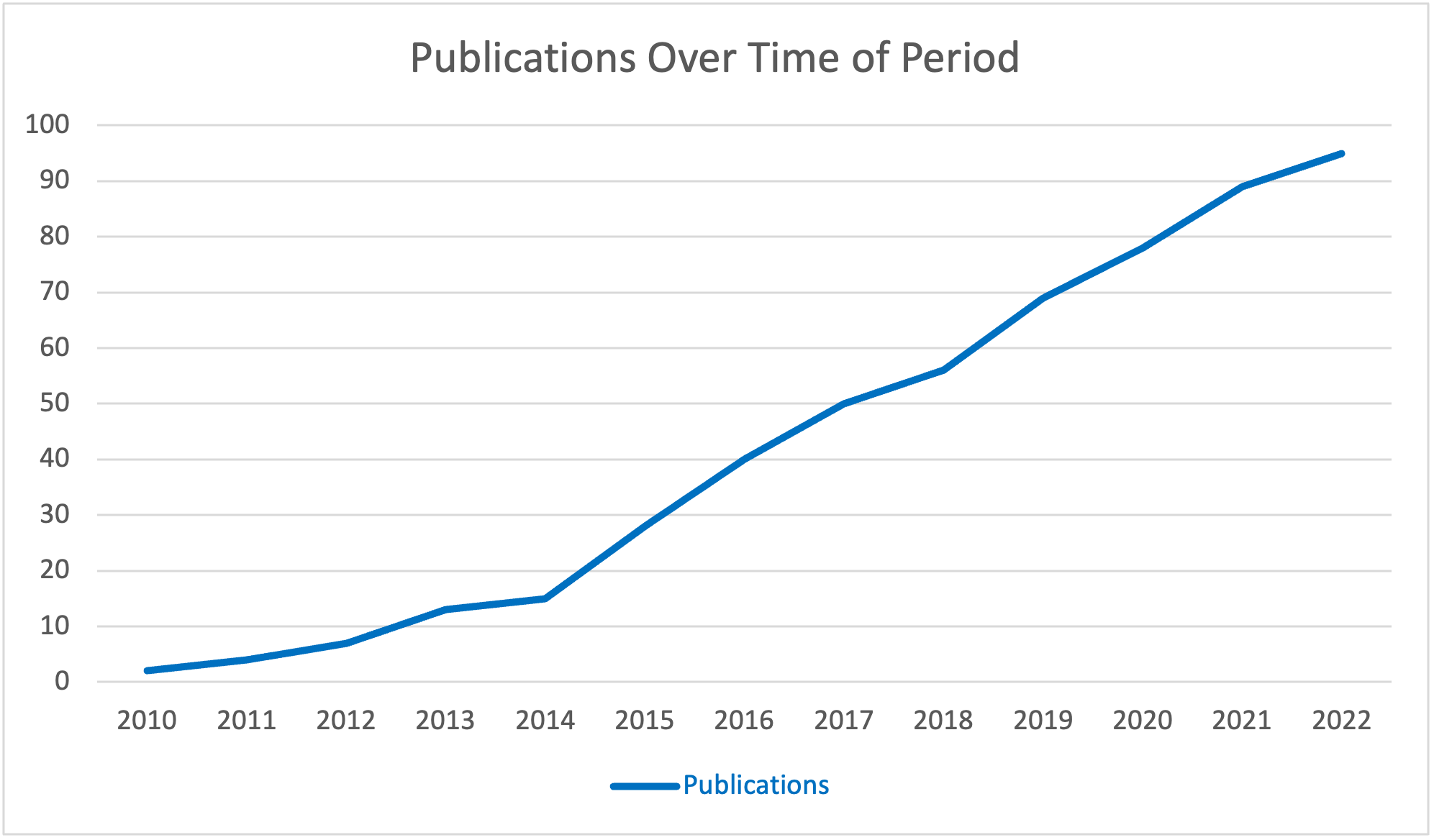}
	    \caption{Number of primary studies published over time.}
	    \label{MyFig3}
\end{figure}

\subsubsection{Significant keyword counts:}
In order to identify the theme of the selected primary studies we have to summarize the keywords count in the table. Using this table we can determine how much a keyword inspires the topic of the paper. From the table, it can be seen that the word "UAV" and "Cyber-Security" were the most frequent, after that the third most frequent word was "Cyber-attack". So, it can be said that lately, everyone is focusing on cyber attacks over UAVs or any other IoT devices.

\section{Findings}
Following a thorough reading of each primary research paper, pertinent qualitative and quantitative information was gathered, and then condensed in Table \ref{Mytable5}, Table \ref{Mytable6} and Table \ref{Mytable7}. With respect to how Unmanned Aerial Vehicles was addressing a specific issue, all of the major studies had a focus or topic. Additionally, Table \ref{Mytable5}, Table \ref{Mytable6} and Table \ref{Mytable7} below lists each paper's focus.
To make it easier to classify the primary studies' topics, each paper's focus was further divided into wider groups.
In the area of the network between UAV and the base station, studies that focused on virtual machines, networking, and virtual network management were gathered. In the area of data storage and sharing, studies with an emphasis on peer-to-peer sharing, encrypted data storage, and search were compiled.
fig 3. demonstrates the percentages of the 30 primary studies' various topics that met the criteria for inclusion in the data analysis.
The themes found in the primary research show that the security and privacy of IoT devices account for nearly half (42 \%) of all studies on applications for cyber security. The second most common subject, at a proportion of 20 \%, is data sharing and storage. Networking for encrypted data and for avoiding tampering with file names and data within is included in the research. The third most prevalent subject, networks, accounts for 16 \% of all themes and is mostly focused on how security and authenticity may be given for UAVs. The fourth most prevalent subject is data privacy and public key infrastructure, each with a proportion of 5 \%.
The fifth most prevalent subject is how Distributed Denial of Service (DDoS) and Internet of Drones (IoD) may successfully host records in a distributed environment to thwart malicious alterations and denial of service assaults. WiFi, GPS, and e-attacks make up the final three prevalent topics on our list, each of which accounts for 4 \%.

\begin{table}[h!]
\caption{Number of keywords used in primary studies.} 
\label{Mytable4}
\setlength{\tabcolsep}{5pt}
\begin{tabular}{p{125pt} p{90pt} }
\hline
\vspace{0.30 mm}Keywords   & \vspace{0.30 mm} Count\\
 \hline
 \vspace{-0.40 mm}UAV   & \vspace{-0.40 mm}3426\\
 \vspace{-1.50 mm} Cyber-Security& \vspace{-1.50 mm} 2897 \\
 \vspace{-1.50 mm}Cyber Attack & \vspace{-1.50 mm}2490\\
 \vspace{-1.50 mm} IoT& \vspace{-1.50 mm} 2289 \\
 \vspace{-1.50 mm}Network   & \vspace{-1.50 mm}2019\\
 \vspace{-1.50 mm} Drone & \vspace{-1.50 mm}  1892  \\
 \vspace{-1.50 mm} Privacy  & \vspace{-1.50 mm}1625\\
 \vspace{-1.50 mm} Application& \vspace{-1.50 mm} 1545  \\
 \vspace{-1.50 mm}Encryption   & \vspace{-1.50 mm}1278\\
 \vspace{-1.5 mm} Wireless Connectivity& \vspace{-1.5 mm}  1190  \\
 \vspace{-1.5 mm} GPS  & \vspace{-1.5 mm}929\\
 \vspace{-1.5 mm} Ground Control& \vspace{-1.5 mm} 857 \\

 \hline
\end{tabular}
\end{table}
\begin{figure}[h]
	    \centering
	    \includegraphics[width=0.9\linewidth,keepaspectratio]{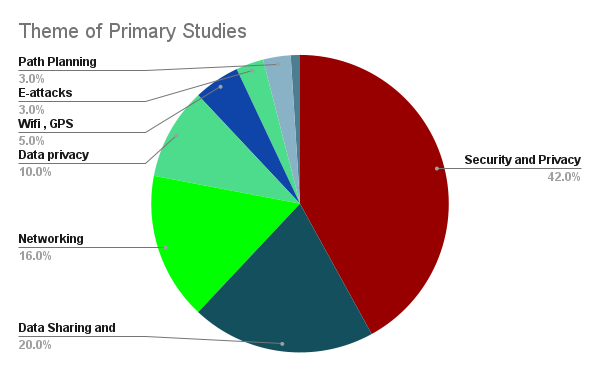}
	    \caption{Themes of the Primary Studies}
	    \label{MyFig4}
\end{figure}

\begin{table}[]
\caption{Main findings and themes of the primary studies.} 
\label{Mytable5}
\centering
\tiny

\begin{tabular}{|c|c|c|}
\hline
\multicolumn{1}{|l|}{Primary Studies} & \begin{tabular}[c]{@{}c@{}}Key Qualitative and Quantitative Data\\ Reported\end{tabular}  & \begin{tabular}[c]{@{}c@{}}Types of Security \\ Applications\end{tabular}  
\\ 
\hline
\cite{bortoff2000path}  & \begin{tabular}[c]{@{}c@{}}A path is created in the first \\ stage by building and searching \\a graph based  on Voronoi polygons.  The graph is used\\ as a starting condition in the\\ second stage, which simulates \\nonlinear ordinary differential equations. The dynamics of a group \\of virtual masses situated\\ in a force field are described \\by the ODEs.\end{tabular} & \begin{tabular}[c]{@{}c@{}} Path planning\\ of UAV \end{tabular}                                                                           \\ \hline
\cite{thakur2015investigation}                                    & \begin{tabular}[c]{@{}c@{}}Any company that lacks comprehensive security\\ measures and security procedures is in significant danger.\\ Cyber crimes are on the rise in the modern period,\\ which has increased the need for system\\ security or even network-wide security.\\\end{tabular}                                                                             &  \begin{tabular}[c]{@{}c@{}} Firm Data\\ Security  \end{tabular}                                                                         \\ \hline
\cite{shakhatreh1805unmanned}                                   & \begin{tabular}[c]{@{}c@{}}The next major advancement in UAV technology\\, known as smart UAVs, promises to open up\\ new possibilities in a variety of applications, particularly\\ in civil infrastructure. Over \$45 billion in UAV utilization\\ is predicted to be dominated by civil infrastructure.\\ We discuss UAV civil applications\\ and their difficulties in this study.\\\end{tabular}                                                                                       & \begin{tabular}[c]{@{}c@{}} UAV as civil\\ infrastructure\end{tabular}                            \\ \hline
\cite{mozaffari2019tutorial}                                    & \begin{tabular}[c]{@{}c@{}}Unmanned aerial vehicles (UAVs) are used\\ increasingly more often. Within a cellular\\ network, UAVs can function as mobile terminals\\ or aerial base stations. Key principles for the\\ analysis, optimization and design of UAV-based wireless\\ communication systems are presented in this lesson.\\\end{tabular}                                                                                                              & \begin{tabular}[c]{@{}c@{}}Cellular\\ network\end{tabular}                            \\ \hline
\cite{waharte2010supporting}                                    & \begin{tabular}[c]{@{}c@{}}Autonomous UAVs can be used in search and rescue efforts\\ to scan the area and gather information about\\ a missing person's whereabouts. Some important factors\\ must be taken into consideration while designing\\ the search algorithms in order to reduce the amount of time it\\ takes to discover the victim. We examine how these \\elements may influence the search task and outline some of the\\ directions for further study.\\\end{tabular}                                            & \begin{tabular}[c]{@{}c@{}}Search\\ Algorithm \end{tabular}                         \\ \hline
\cite{article1} & \begin{tabular}[c]{@{}c@{}}Drone technology has new applications \\ because of the\\ Internet of Things (IoT), but it also faces new risks due\\ to its quick development. Small drones have design problems\\ as well as security and safety definition problems.\\ These little drones still need a design that is developed\\ enough to meet domain criteria.\\\end{tabular} & Issues in IoT applications                                                                          \\ \hline
\cite{article} & \begin{tabular}[c]{@{}c@{}}Drone security and privacy issues are\\ becoming more and more of a worry.\\ This essay offers a succinct explanation of\\ the privacy and security issues surrounding\\ unmanned aerial vehicles. Additionally, it talks\\ about related constraints, vulnerabilities, and\\ current defenses against various assaults.\\ The study ends with a discussion\\ of open research areas and suggestions\\ for how to deal with these issues.\\\end{tabular}                                                                    & \begin{tabular}[c]{@{}c@{}}Security and \\Privacy issue\end{tabular}                 \\ \hline
\cite{hartmann2013vulnerability} & \begin{tabular}[c]{@{}c@{}}Provide the first approach to a UAV-specific\\ risk assessment based on the provided \\services and communication infrastructures.\\\end{tabular}                                    & \begin{tabular}[c]{@{}c@{}}Risk \\Assessment\end{tabular}                    \\ \hline
\cite{gudla2018defense} & \begin{tabular}[c]{@{}c@{}}UAVs carry, collect, or communicate sensitive\\ information which becomes a target for\\ cyber-attacks. Securing the communication \\network between the operator and the\\ UAV is crucial. MTD technique\\ changes the static nature of the systems\\ to increase the difficulty and cost\\ of mounting attacks.\\\end{tabular}                                                             & \begin{tabular}[c]{@{}c@{}}MTD \\Techniques\end{tabular}                        \\ \hline
\cite{basan2021self} & \begin{tabular}[c]{@{}c@{}}Global Navigation Satellite System (GNSS) is\\ widely used for locating drones.\\ This is because of the simplicity\\ and low cost of this technology.\\ There are many security threats to\\ GPS navigation. These are primarily\\ related to the nature of the GPS\\ signal. We discuss methods of protection\\ against this type of attack.\\\end{tabular}                                                                                                                                & \begin{tabular}[c]{@{}c@{}}Safe \\navigation\end{tabular}                        \\ \hline
\cite{tsao2022survey}  & \begin{tabular}[c]{@{}c@{}}UAV systems may collaborate with one\\ another to carry out a variety\\ of functions. UAVs can be used for\\ business purposes including transportation of commodities,\\ military surveillance, and search and rescue\\ operations. Such attacks have become more\\ frequent, and they can have terrible\\ consequences. Security measures for UAV systems\\ and networks are being investigated by industry\\ and standardization organizations. Our investigation focuses\\ on risks to the Internet of\\ Drones architecture as well as security\\ and privacy concerns for UAVs\\ while constructing flying ad-hoc networks (FANETs).\\\end{tabular}                                                                                     & Ad-Hoc Network                                                                                \\ \hline
\cite{KONERT2021292} & \begin{tabular}[c]{@{}c@{}}Paper will analyze the legal and ethical\\ issues on using drones as a weapon.\\ Unmanned aerial vehicles, (UAVs), \\called drones, have recently increased their\\ role from simple surveillance and reconnaissance\\ to increasingly controversial targeted killings. Autonomy\\ refers to respect for human\\ autonomy (in contrast with the autonomy\\ of a drone) and includes the\\ free choice of individuals and groups.\\\end{tabular}                                                                          & \begin{tabular}[c]{@{}c@{}}Ethical\\ Issues\end{tabular}                  \\ \hline
\cite{yahuza2021internet}  & \begin{tabular}[c]{@{}c@{}}Drones' access to regulated airspace is\\ linked via the Internet of\\ Drones (IoD), a decentralized network and management\\ architecture that also offers inter-location\\ navigation services.\\ All security and privacy risks that\\ are present for IoT networks also\\ exist for the IoD network.\\ The total impact of the\\ IoD paradigm has been considerably\\ constrained by security and privacy\\ concerns. This essay tries\\ to evaluate current trends in network\\ security, privacy, and\\ data protection concerns. Then, we\\ emphasize the demand for a secure\\ IoD architecture and create one.\\\end{tabular}                                                                                          & \begin{tabular}[c]{@{}c@{}} IoD and\\ Privacy Threat  \end{tabular}                                                                               \\ \hline

\end{tabular}
\end{table}

\begin{table}[]
\caption{Main findings and themes of the primary studies.} 
\label{Mytable6}
\centering
\tiny

\begin{tabular}{|c|c|c|}
\hline
\multicolumn{1}{|l|}{Primary Studies} & \begin{tabular}[c]{@{}c@{}}Key Qualitative and Quantitative Data\\ Reported\end{tabular}                                                                                                                                                                    & \begin{tabular}[c]{@{}c@{}}Types of Security \\ Applications\end{tabular}          \\ \hline
\cite{albalawi2019data} & \begin{tabular}[c]{@{}c@{}}Especially DDoS and Distributed Denial\\ of Service (DDoS) assaults, the Internet\\ of Flying Things (IoT) is susceptible to cyberattacks.\\ In order to remedy the security\\ flaw, this paper suggests a deep\\ learning system based on experience that\\ can handle DoS, D-DoS, and\\ ping-of-death assaults. The suggested plan\\ makes use of the idea of an\\ intrusion detection system (IDS).\\\end{tabular}    & \begin{tabular}[c]{@{}c@{}}DDoS\end{tabular}                  \\ \hline
\cite{lagkas2018uav} & \begin{tabular}[c]{@{}c@{}}Unmanned aerial vehicles (UAVs) have the\\ potential to revolutionize a wide\\ range of industries, including the military,\\ security, healthcare, surveillance, and traffic\\ monitoring. UAVs with cameras, sensors,\\ and GPS receivers have a lot of\\ promise when using emerging technologies like\\ 4G/5G networks. Before UAVs may be\\ used effectively, a number of\\ problems, including those relating\\ to administration, security, and privacy\\ must be overcome.\\\end{tabular}                                                                   &  \begin{tabular}[c]{@{}c@{}}Traffic Monitoring\\ and GPS  \end{tabular}                                                                            \\ \hline
\cite{siddappaji2020role} & \begin{tabular}[c]{@{}c@{}}Unmanned aerial vehicles, or drones,\\ are examples of cyber-physical\\ systems (CPS), which combine physical, networking,\\ and computational processes. Systems used\\ by drones to fly are dependent\\ on embedded computing power and virtual\\ cyber networks to function.\\ Because of the drone's distinctive network\\ and distributed physical systems that are\\ situated in far-off locations,\\ it has been determined that they\\ are susceptible to cyberattacks.\\\end{tabular}                                                                             & \begin{tabular}[c]{@{}c@{}}Cyber-Physical\\ Systems.\end{tabular}                    \\ \hline
\cite{hooper2016securing}& \begin{tabular}[c]{@{}c@{}}It is possible to take advantage\\ of flaws in the Wi-Fi\\ access point utilized by the Parrot\\ Bebop UAV and the usual ARDiscovery\\ Connection procedure. The UAV's rotors\\ might be catastrophically and instantly\\ disabled in midflight if these vulnerabilities\\ are exploited. We claim that Wi-fi-based\\ commercial UAVs need a robust\\ security architecture that employs a defense-in-depth\\ strategy based on the literature\\ and our own penetration tests.\\\end{tabular}                                                                      & \begin{tabular}[c]{@{}c@{}}WiFi\\ vulnerabilities     \end{tabular}                                                                           \\ \hline
\cite{parikh2017cyber} & \begin{tabular}[c]{@{}c@{}}The assault, danger, and vulnerabilities of\\ cyberinfrastructure—including networks, business\\ networks, and intranets—are examined\\ in this article. It goes through\\ the significance of network\\ intrusions, cybercrime, and the causes of\\ its rapid expansion. The research comes\\ to the conclusion that while technology\\ can help lessen the effects of\\ cyberattacks, people are nonetheless vulnerable\\ because of their actions and\\ psychological tendencies.\\\end{tabular}                                                                                                                                                     & \begin{tabular}[c]{@{}c@{}} Cyber\\ issues   \end{tabular}                                                                        \\ \hline
\cite{cremer2022cyber} & \begin{tabular}[c]{@{}c@{}}Cybercrime is estimated to have cost\\ the global economy just under\\ USD 1 trillion in 2020. The average\\ cyber insurance claim will rise\\ from USD 145,000 in 2019 to\\ USD 359,000 by 2020. The lack\\ of available data on cyber\\ risk poses a serious problem for\\ stakeholders seeking to tackle this issue.\\ We posit that the lack of\\ available data on cyber risk poses\\ a serious problem for stakeholders\\ seeking to tackle this issue.\\\end{tabular}                                                               & \begin{tabular}[c]{@{}c@{}} Cyber data\\ crime      \end{tabular}                                                                          \\ \hline
\cite{almulhem2020threat} & \begin{tabular}[c]{@{}c@{}}Threat modeling can be extremely\\ important in realizing the idea of\\ "secure by design" for Multi-UAV\\ systems, which are still in the early\\ phases of design. In order\\ to investigate and identify\\ dangers affecting the Internet of Drones (IoD) \\architecture, we use a threat modeling\\ approach known as threat trees in\\ this article.\\\end{tabular}                                                                  & \begin{tabular}[c]{@{}c@{}} Threat\\ Modelling\end{tabular}                         \\ \hline
\cite{yoon2017security} & \begin{tabular}[c]{@{}c@{}}Although there are many applications\\ for UAVs, security has always been\\ a top priority. UAV transports\\ private information that is sensitive and\\ poses serious risks if unscrupulous attackers\\ utilize it to their advantage.\\ This study suggests a method for\\ preventing unidentified attackers from\\ controlling commercial UAV hardware\\ or network channels. In the article,\\ it was suggested that a Raspberry\\ Pi-based DoS assault, an extra encrypted \\communication channel, and authentication algorithms be\\ used to keep control of the UAV\\ in a hijacking situation.\\\end{tabular}                                                                                                                                     & \begin{tabular}[c]{@{}c@{}}Authen-tication\\ algorithm\end{tabular}                        \\ \hline
\end{tabular}
\end{table}

\begin{table}[]
\caption{Main findings and themes of the primary studies.} 
\label{Mytable7}
\centering
\tiny

\begin{tabular}{|c|c|c|}
\hline
\multicolumn{1}{|l|}{Primary Studies} & \begin{tabular}[c]{@{}c@{}}Key Qualitative and Quantitative Data\\ Reported\end{tabular}  & \begin{tabular}[c]{@{}c@{}}Types of Security \\ Applications\end{tabular}  
\\ 
\hline

\cite{challita2019machine} & \begin{tabular}[c]{@{}c@{}}Numerous obstacles must be overcome\\ in order to provide UAVs  with the \\secure operation and dependable  wireless communication.\\ These encompass \\ authentication, mobility management and \\ handover, cyber-physical threats, and \\interference management. Such problems \\are addressed using\\ ANN-based solution strategies.\\ The strategies that have been \\ presented allow\\ UAVs to utilize wireless system\\ resources in a flexible manner \\ while yet maintaining \\real-time operational security.\\\end{tabular}                                                                                                                                                   & \begin{tabular}[c]{@{}c@{}}ML for \\ Wireless\\ Connectivity \end{tabular}                        \\ \hline

\cite{qiao2017vision} & \begin{tabular}[c]{@{}c@{}}Networked virtual machine security settings\\ that use private block-chain;\\ IBM's Hyper-ledger Fabric served as an\\ example. sufficient characteristics\\ to enable the researchers' suggestions\\\end{tabular}                                                    & \begin{tabular}[c]{@{}c@{}}Block-chain\end{tabular}                            \\ \hline
\cite{jensen2019blockchain} & \begin{tabular}[c]{@{}c@{}}The GPS signal is weakly received\\ by devices on the ground\\ and is open and unencrypted. GPS\\ transmissions are hence susceptible\\ to jamming and spoofing. UAVs might go\\ out of control or even be\\ hijacked as a result of GPS spoofing.\\ Our approach was tested on\\ the DJI Phantom 4 UAV.\\\end{tabular}                                                          & \begin{tabular}[c]{@{}c@{}}GPS based on\\ spoofing\end{tabular}                           \\ \hline
\cite{fei2018cross} & \begin{tabular}[c]{@{}c@{}}Unmanned aerial vehicles are a\\ fast-evolving cyber-physical platform, and\\ as their capabilities and use improve,\\ so do the security risks\\ they face. In this study,\\ we want to create a general security\\ framework called Blue Box that can\\ recognize and respond to such attacks.\\ The results of the experiments\\ proved that Blue-Box was capable\\ of both identifying different\\ cyber-physical threats and offering\\ a way to recover from them.\\\end{tabular}     & \begin{tabular}[c]{@{}c@{}}Method against\\ cyber attack\end{tabular}                        \\ \hline
\cite{mansfield2013unmanned} & \begin{tabular}[c]{@{}c@{}}Insecure mobile and wireless networks,\\ as well as smart gadgets, provide\\ a risk of unintended data sharing\\ and UAV control loss to adversaries.\\ When it comes to identifying cyber\\ security risks and ensuring the right\\ security countermeasures are in place,\\ the Department of Defense has failed\\ to create a threat model and\\ risk assessment. This essay will examine\\ the hardware, software, and communication\\ channels that make up smart devices'\\ cyber security vulnerabilities.\\\end{tabular}                                                                                              & \begin{tabular}[c]{@{}c@{}}Ground control\\ of UAV\end{tabular}                        \\ \hline
\cite{chen2018privacy} & \begin{tabular}[c]{@{}c@{}}An open-source UAV simulator called\\ Air Sim contains characteristics including\\ simplicity in construction, effective motion\\ capture, effective collision and\\ obstacle recognition, and physics models.\\ IoD communication frameworks must be\\ created securely while minimizing\\ performance compromises due to the huge\\ volume of data that is currently\\ being sent between IoD devices.\\\end{tabular}                                    &  \begin{tabular}[c]{@{}c@{}}Open Air\\ space issue    \end{tabular}                                                                           \\ \hline
\cite{inproceedings} & \begin{tabular}[c]{@{}c@{}}Concerns about electronic attacks,\\ hacking, and hijacking on civilian\\ UAVs are all made much more complicated\\ by the lack of attribution and\\ security procedures to prevent these\\ actions. The public's careless usage\\ of drones raises a variety of problems,\\ from privacy infractions to potentially\\ very dangerous circumstances like failing\\ to avoid regulated airspace.\\\end{tabular}                                                                             & \begin{tabular}[c]{@{}c@{}}Electro-nics\\ Attacks\end{tabular}                            \\ \hline
\cite{iqbal2021study} & \begin{tabular}[c]{@{}c@{}}Many UAV-related security incidents are\\ reported nowadays. Governments around\\ the world have started to regulate the\\ use of UAVs. In this paper,\\ we investigate the security of existing\\ operating systems used in consumer\\ and commercial UAVs. We discuss\\ several research challenges for developing\\ a secure operating system for UAVs.\\\end{tabular}              &   \begin{tabular}[c]{@{}c@{}}Operating System\\ Security       \end{tabular}                                                                         \\ \hline
\cite{abbaspour2016detection} & \begin{tabular}[c]{@{}c@{}}UAV control systems should be built\\ to be as secure and resistant\\ to assaults as feasible, including\\ fault data injection (FDI) attacks.\\ In this research, a novel technique\\ is presented to identify\\ malicious flaws and cyber-attacks in UAVs.\\ To find the inserted errors\\ in a UAV's sensor, an adaptive\\ neural network is employed.\\\end{tabular} & \begin{tabular}[c]{@{}c@{}}Fault Data\\ Injection\end{tabular} \\ \hline

\end{tabular}
\end{table}

Below are discussed in-depth the three research questions for better understanding and information for researchers gathered from the several primary studies listed above.

\subsubsection{RQ1:  What are recent advances for preventing information loss in UAVs?}
It is necessary to answer the research question after the primary studies have been selected. With that in mind, it can be said that the most recent advancement in the prevention of information loss in UAVs is complete knowledge of cyber attacks. 
At the current state, researchers have reviewed all primary studies and determined that all the possible attacks that can happen over an Unmanned Aerial Vehicle is classified into 2 different part and they are discussed as follows:

\begin{enumerate}
    \item Physical - Potential eavesdropping is the most of attack that can happen over unmanned aerial vehicles \cite{yahuza2021internet}, \cite{siddappaji2020role}, \cite{hooper2016securing}. The second most attack is traffic analysis \cite{tsao2022survey},\cite{lagkas2018uav} followed by Interference  \cite{almulhem2020threat}, \cite{yoon2017security}
    \item Electronic - this attack includes scanning, Man-in-the-middle, WiFi-jamming \cite{abbaspour2016detection}, \cite{iqbal2021study}, \cite{qiao2017vision}. There are additional other attacks too such as, denial of service, Password Breaking, and Reconnaissance. \cite{thakur2015investigation} \cite{abbaspour2016detection}, \cite{mozaffari2019tutorial}, \cite{article},\cite{hooper2016securing},\cite{cremer2022cyber}, \cite{yoon2017security}. 
\end{enumerate}

\subsubsection{RQ2: Is any secured method available for communications of the drones?}
From our primary research, Today, UAVs have advanced significantly in both the military and defense sectors, such as security missions involving reconnaissance, surveillance, and monitoring of the environment, as well as in the civil sectors, including urban planning, search and rescue, law enforcement, traffic monitoring, accident management, agricultural assessment, and entertainment.
environmental surveillance Rapid growth is being seen in photography, infrastructure monitoring, and rescue operations.
\\
So, in both use cases, military and civil use, different communication system is used in drones. In military usage, the communication protocols have to be secured, because if a communication breach occurs, many important military security-related data might be compromised. Whilst, In Civil Sectors, more focus is on data leaks rather than communications.
There are several methods of communication in drones and they are discussed below:
Most UAV communication is based on three types and they are as below: 

\textbf{Radio Frequency} - As we know, all communication for UAVs is wireless. So, the controller sends a radio signal of the next action for the drone. The radio frequency has to be between 2.4 GHz and 5.8 GHz as it is a normal frequency bandwidth in which UAV communications work. \cite{mansfield2013unmanned},\cite{challita2019machine}, \cite{yoon2017security}, \cite{albalawi2019data}

\textbf{WiFi Controls} -     WiFi-enabled drones are typically used to stream footage to a tablet, PC, or smartphone. These gadgets can also be used by a controller to remotely control the drone, \cite{hooper2016securing},\cite{albalawi2019data}. For this type of communication, first, one WiFi network has to be created and deployed, so that tests over drones can take place. There are several WiFi nodes in this Aerial WiFi network, the connectivity of drones jumps from one node to another and that brings us a continuous connection to the UAV devices. However, it is not efficient as the Ad-hoc method.\cite{albalawi2019data},\cite{tsao2022survey},\cite{article1}

\textbf{GPS} - A GPS drone is equipped with a GPS module that allows it to know its position relative to a network of orbiting satellites. By connecting to signals from these satellites, drones can perform functions such as position-keeping, autonomous flight, homecoming, and waypoint navigation, \cite{lagkas2018uav}. There are different factors that affect the use of this communication type, they are higher cost, battery drainage, GPS-Denied Scenario, and others.  \cite{lagkas2018uav},\cite{qiao2017vision}, \cite{fei2018cross}

\subsubsection{RQ3: How is the security measured for overtaking attacks done at various types of stages like level of communication, hardware level, sensor level, and software level?}
As described in the research question itself the measurement for preventing the cyber attack happens over several stages, any vulnerable leaks in each stage can result in overtaking control of drones.
Each stage is described below:

\begin{itemize}

 \item {\textbf{Communication Level}} - At the communication level, attacks that could happen are eavesdropping, network collision, and others. To prevent this advanced Machine Learning algorithms are used to secure communication. GPS is suggested to use for communication, as it is worldwide and it communicates through satellite. \cite{KONERT2021292}

    \item {\textbf{Hardware Level}} - We can consider this as ground level, from where all the communications and orders are relayed to UAVs. For this level, a multi-factor authentication system is established to secure communication. In addition, strong and periodic passwords are used. The hash key can be used to secure the password.\cite{yoon2017security}

 \item {\textbf{Sensor Level}} -     Mostly 4 types of sensors are being used in UAVs, Optical Cameras, Ground Penetrating Radar, Lightweight Portable Radiometer, and Lidar (Light Detection and Ranging) sensors. To protect this sensor, some changes to hardware equipment are required. We can use an extra layer of protected glass over a sensor to protect it from potential threat. As per the logical threats, a multi-factor authentication system and a Hash-key password system are being used. For the ground Penetration radar, encryption of the frequency is required to secure the communication. \cite{qiao2017vision}, \cite{fei2018cross}

 \item {\textbf{Software level }}- At the software level, many attacks can be done using MALWARE. Protection against blocking by calculating position by drift measurements using odometer techniques. We can use blockchain and Machine Learning algorithms to protect the software that communicates with the drone. The drone collects all kinds of data text, images, location, and videos. To transfer this data there is software that works behind this. So, we can embed that software with security levels to prevent it from leaking data.\cite{inproceedings},\cite{abbaspour2016detection},\cite{thakur2015investigation}

\end{itemize}

\section{Discussion}
There is a considerable number of research papers available for the keyword UAV or Unmanned Aerial Vehicle. The technologies for UAV and multi-UAV systems have been developed in the past 7 years and can be told as the beginning level of development in the field.

The majority of those listed in primary research are experimental hypotheses or notions with limited quantitative information and few real-world implications for the challenges of today.
Many of the more useful security options provided in the  subsequent primary studies demonstrate creative methods for addressing a variety of issues relating to data security, mutability, and networking.
The solutions delivered in the studies highly depend on the number of the system environments and surrounding and various factors like weather, labor, and attacks. Due to the reasons mentioned \cite{chen2018privacy} open-air space, and practical experiments of a certain length  have shown the effectiveness of UAV applications in today's world over conventional security. Notable work in \cite{thakur2015investigation} is seen for data security to secure data networking and communication. This study discovers that while creativity has a role to play in decreasing the effect of digital assaults, threats, and weaknesses, cyber security threats and approaches have a role to play in minimizing the impact of digital attacks, risk, and vulnerability. Although cyberattacks can be mitigated, there does not currently seem to be a conclusive way to defeat such network security risks.

The researchers used the current \cite{article} network security methods like ML-based IDS, which is considered to be a successful implementation. There are various IDS like rule-based, signature-based, and anomaly-based.  This study finds that while cyber security threats and techniques have a role to play in limiting the impact of digital attacks, risk, and vulnerability, creativity also has a role to play in reducing the effect of digital assaults, danger, and weakness.\cite{abbaspour2016detection} Although assaults can be lessened, there doesn't appear to be a definitive solution to eliminate such network security vulnerabilities at the moment. Both performance measurements and security aspects (authentication, forward secrecy, reverse secrecy, etc.) have been considered in the analysis of the frameworks (communication cost, computation cost, storage cost, signature generation time, signature verification time, etc.). The review study concludes with difficulties and recommendations for the foreseeable future for the researchers.

\cite{chen2018privacy}. The fault data injection \cite{abbaspour2016detection}  helps the data to be safe in the electronic attack \cite{inproceedings} which is like privacy infractions to circumstances in the airspace.\cite{jensen2019blockchain} Regular network structure swapping makes it challenging to implement the flying-AntHocNet routing control scheme, which was inspired by a systematic environment-based approach and exhibits optimal simulation results for metrics like end-to-end delay, packet loss\cite{thakur2015investigation}, data-packet drop-count, and throughput analysis in comparison to traditional routing techniques like DSDV, DSR, AOMDV, M-DART, and Z-R-P, which are introduced in aerial networks\cite{waharte2010supporting}. In order to protect society from cyberattacks, smart cities are technologically implementing internet-of-everything abstraction.
\section{Future Research Directions of UAV Cyber Security}
Based on the outcomes of this survey along with our own observations, we suggest the following blockchain research areas that need further study:
\subsection{UAV: Deep and Machine Learning:}
\cite{lee2017optimal} Recently, a lot of support has been given to machine learning and deep learning techniques in a variety of UAV-related applications, including allocation of resources, obstacle detection, tracking, trajectory planning, and battery scheduling. In order to complete the help device accurately and without the risk of collision, it will be necessary to design nano UAVs that are much smaller, lighter, and smarter than existing UAV models.\cite{noorwali2021efficient} This will be facilitated by the creation of more accurate algorithms and an increase in onboard computational power. Furthermore, reliable data availability can help UAVs execute precise control, path planning, and vision jobs. 
\\
The mission route of UAVs should be optimized while the energy usage of flights with emergency braking is reduced by using better tracking and path planning algorithms. \cite{shakeri2019design}While path planning of complicated paths to avoid obstacles and identify the shortest path to spend minimal energy may be carried out by employing a multi-objective optimization algorithm, recent work on the issue of tracking is mostly based on heuristic techniques.\cite{carrio2017review} The review included a number of deep-learning applications for UAVs, including feature extraction. The UAV may be equipped with numerous cameras to capture a range of picture types for further processing. As the UAV searches the surroundings for an appropriate path, planning for path motion, tracking, and manipulation were given. Deep learning-based UAV motion control is another use.
\subsection{UAV: Security and Privacy}
Due to their cellular networking and limited processing capabilities, UAV security and safety are crucial factors.\cite{lee2017optimal} They might be threatened by intrusion attempts, which would jeopardize the acquired data's security and privacy. It may be replaced or stolen. New onboard methods are thus needed to ensure privacy during the voyage. Recent technologies like blockchain technology and physical layer protection need further study and development to reach the necessary security level with the necessary quality and dependability\cite{carrio2017review}.
\\
The effectiveness of optical wireless communications (OWCs) in 4th Generation or, 5th Generation, and beyond 5G mobile networks has been established.\cite{oubbati2020softwarization}They are frequently utilized in UAV communication, and the 6G mobile network is anticipated to employ them. However, a number of issues with this technology still need to be resolved, including the high likelihood of signal blocking, power consumption, and weather conditions.
There are some practical issues that need to be answered or future investigations and understandings like:

\begin{enumerate}
      \item What development can be made for high end-to-end delay for non-delay tolerant UAVs when the link between the control and data panel is affected?
      \item  While being energy-efficient UAVs how will it make sure about the confidentiality of the data?
\end{enumerate}

\subsection{UAV: Detection System for Intrusion}
Real-time network traffic analysis is needed to find intrusions against UAVs while they are in flight. In order to do this, employing an IDS for Drones allows the detection of many intrusion types, including assaults that alter signals, malware, attack routing, and message forging \cite{choudhary2018intrusion}. Anomaly detection framework development to track malicious actions also plays a significant part in identifying attack patterns \cite{schumann2015r2u2}. Additionally, the use of honey pots and honey nets in conjunction with IDS can aid in shielding the flying mission from nefarious actors \cite{franco2021survey}.
\\
However, having additional sources of information can also result in higher communication costs and higher overhead costs for processing. Creating such solutions is difficult since security and performance are currently tradeoffs. To monitor UAV communications and identify threats, lightweight IDSs must be put into use. In this regard, several systems make use of the flight's behavioral profile to find unusual activity and malicious intrusions \cite{birnbaum2015unmanned}. However, such methods cannot identify cyber assaults that compromise UAVs while guaranteeing that the flying pattern remains constant.

\section{Conclusion}\label{conc}
This study has uncovered previous studies that examine how Unmanned Aerial Vehicle solutions might exacerbate cyber security issues. This study examined a UAV system's overall security hazard. The thorough threat analysis of the system is intended to assist both the system's designers and users in recognizing potential vulnerabilities and putting appropriate mitigation and recovery mechanisms in place. It is still difficult to pinpoint which risks could have the greatest impact on UAV systems because the majority of the information about the current safeguards is classified. Working on a few of these dangers and utilizing mission data to better precisely simulate them will be part of future development.
\subsection{Potential research agenda 1: Fuzzing UAV}
To accomplish the goal of a wide coverage and depth test, the fuzzing is guided by the information gathered about the program execution. The technological issues brought on by the uniqueness of UAV systems, such as the in-applicability of standard fuzzing testing methodologies, secret communication protocols, and difficulties in monitoring the running state must be resolved. An innovation would be to accomplish generation-based fuzzing by extracting syntactic information from the proprietary protocols of UAV systems.
\\
Another area that merits investigation in order to help grey-box fuzzing is UAV simulation technology. One may properly extract execution information from simulation systems by using memory activity detection, program stack status, and hardware output to direct the process of fuzzing.
\subsection{Potential research agenda 2: BlockChain}
Blockchain is a brand-new way to use cloud computing including point-to-point transmission, consensus, encryption, and distributed data storage. Attackers find it challenging to change or remove the records that may be used to record, gather, and search information about UAV systems, such as where the UAV has been or what it has done, because of its decentralized and encrypted communication capabilities, notably the redundancy check of data. It may be used to spot unauthorized UAV systems or unusual operating modes, making it more challenging for attackers to get data from UAVs or alter flight plans. Although it is unclear what effect blockchain will have on UAV security, it will be a future technology and development pattern to be taken into account.

\bibliographystyle{IEEEtran}
\bibliography{reference}

\end{document}